\documentclass[pra,twocolumn,aps,showpacs,tightenlines,superscriptaddress]{revtex4}
\usepackage{amssymb,amsmath,amsthm,epsfig,color,graphicx,times}

\usepackage{hyperref}
\usepackage{enumerate}

\begin{document}

\title{Mutual information and spontaneous symmetry breaking}
\author{A. Hamma}
\thanks{Corresponding author: hamma.alioscia@gmail.com}
\affiliation{Center for Quantum Information, Institute for Interdisciplinary Information
Sciences, Tsinghua University, Beijing 100084, P.R. China}

\author{S. M. Giampaolo}
\affiliation{Dipartimento di Ingegneria Industriale, Universit\`a degli Studi di Salerno, I-84084 Fisciano (SA), Italy}

\author{F. Illuminati}
\affiliation{Dipartimento di Ingegneria Industriale, Universit\`a degli Studi di Salerno, I-84084 Fisciano (SA), Italy}

\date{October 29, 2015}

\begin{abstract}
We show that the metastable, symmetry-breaking ground states of quantum many-body Hamiltonians have vanishing quantum mutual information
between macroscopically separated regions, and are thus the most classical ones among all possible quantum ground states. This statement
is obvious only when the symmetry-breaking ground states are simple product states, e.g. at the factorization point. On the other hand,
symmetry-breaking states are in general entangled along the entire ordered phase, and to show that they actually feature the least
macroscopic correlations compared to their symmetric superpositions is highly non trivial. We prove this result in general, by
considering the quantum mutual information based on the $2-$R\'enyi entanglement entropy and using a locality result stemming from
quasi-adiabatic continuation. Moreover, in the paradigmatic case of the exactly solvable one-dimensional quantum $XY$ model, we further
verify the general result by considering also the quantum mutual information based on the von Neumann entanglement entropy.
\end{abstract}


\maketitle

\section{Introduction}

The emergence of a macroscopic classical behavior from a microscopic quantum world can be explained in terms of decoherence to the
environment that quickly destroys the coherent superpositions of macroscopic objects (Schr\"odinger cats)~\cite{Zurek2003}. The selected
pointer states must then be factorized states with respect to a tensor product structure that is local in real
space~\cite{Zurek1993,ZukerHabib1993}. Similarly, superselection induced by decoherence due to weak interactions with the environment
plays a key role also in the phenomenon of spontaneous symmetry breaking, where different ordered sectors with broken symmetry are
dynamically disconnected and are thus the only states that are metastable~\cite{vanWezel2008}, whence the  notion of spontaneous symmetry
breaking~\cite{Goldstone1962}.

In the paradigmatic case of the quantum Ising model, the ground space of the ferromagnetic phase at zero transverse field $h$ is spanned
by two orthogonal product states $|0\rangle^{\otimes N}$ and $|1\rangle^{\otimes N}$ which are in the same class of pointer states of
the typical decoherence argument, while the symmetric states $\Psi_{\pm} = 1/\sqrt{2}(|0\rangle^{\otimes N} \pm |1\rangle^{\otimes N})$ realize
macroscopic coherent superpositions that are not stable under decoherence~\cite{Zurek2003,vanWezel2008}. Therefore, at zero transverse field $h$,
the situation is very clear: the only stable states are those that maximally break the symmetry of the Hamiltonian, and at the same time, 
those that feature vanishing macroscopic total correlations, including entanglement, between spatially separated regions.
\begin{figure}[t]
\includegraphics[width=7cm]{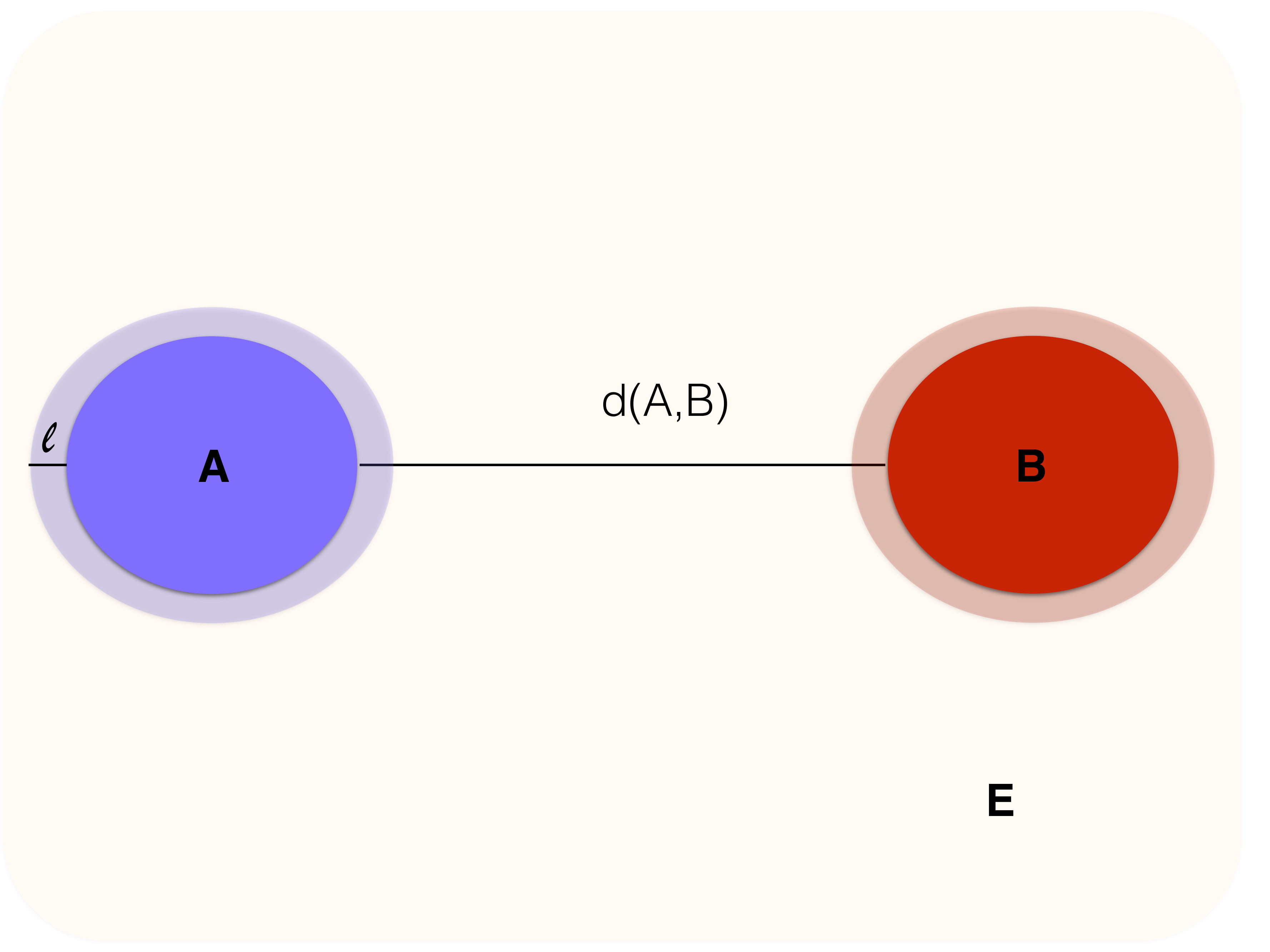}
\caption{A many-body quantum system is partitioned in three distinguishable subsystems, $A$, $B$, and the remainder $E$, so that the total 
Hilbert space acquires the tensor product structure $\mathcal H = \mathcal H_A \otimes \mathcal H_B \otimes\mathcal H_E$. The quantity $d(A,B)$ is the
distance between the two regions $A,B$, and $l$ is the distance defining the new effective support after an adiabatic deformation of
operators with initial support on $\mathcal H_A, \mathcal H_B$.}
\label{system}
\end{figure}

As we turn on the external field $h$, we have a whole range of values where, before a critical value $h=h_c$ is reached, there is a
magnetic order associated to spontaneous symmetry breaking~\cite{BM1971}, and the decoherence argument applies within the entire ordered
phase. This means that, again, the only stable states are those that maximally break the Hamiltonian symmetry~\cite{Cianciaruso2014,ssbhalvorson}.
However, now the symmetry-breaking states are entangled, and their mixed-state reductions on arbitrary subsystems possess in general
nonvanishing entanglement~\cite{Osborne2002,Osterloh2002,Amico2008}, as well as quantum~\cite{TRHA2013,Campbell2013,Cianciaruso2014-1}
and classical correlations~\cite{BM1971}. Indeed, the symmetry-breaking ground states can be, ``locally'', more entangled than some
nearby symmetric states~\cite{Osterloh2006}. On the other hand, it is always implicitly assumed that such states are not
{\em macroscopically} correlated, while their symmetric superpositions are, in complete analogy with the case $h=0$. Although this is
a very plausible picture, a rigorous proof has never been provided, due to the mathematical difficulties in dealing with measures of entanglement
and correlations based on the von Neumann entropy; see, e.g., the difficulties in proving the boundary ({\em area}) law in generic gapped
systems~\cite{arealaw1, arealaw2}, or in proving the stability of topological entanglement entropy in topologically ordered states\cite{TE}. The symmetry-breaking states obey the boundary law for
entanglement~\cite{Holzhey1994,Latorre2003,Vidal2003,Korepin2004}, while the macroscopic correlations featured by the superposition of
two different symmetry broken sectors are of order one. 

The question is then about which quantity one should look at in order to distinguish the presence of macroscopic correlations, among all  
possible source of entanglement and correlations?
Historically, the key concepts that have been considered are the off diagonal long range order (ODLRO)~\cite{odlro} and, more recently, the two-site concurrence (entanglement of formation) at large distance~\cite{vedralnjp,Heaney2007}. If there is either ODLRO or
nonvanishing concurrence between two sites in different clusters $A$ and $B$, then also the two clusters must be entangled, since the
total state of the global system is pure. This is an important point, because the reduced subsystems being in a separable state does not imply that 
there must be no entanglement between the two clusters in a pure state. Even if all the remaining correlations are classical, they are due to the fact that the overall state is a pure entangled state.

On the other hand, the reverse argument needs not apply: it is possible that macroscopic clusters are entangled even if measures of two-point correlations,
like the concurrence, are vanishing. For example, this can happen in the 2D toric code~\cite{topent1} or in the one dimensional cluster
models~\cite{cluster1,cluster2,cluster3} where all two-site concurrences are identically zero and yet the macroscopic block entanglement entropy is finite. Also states that possess volume law for the bipartite entanglement typically have no two-site
entanglement, because of monogamy. Moreover, concurrence and ODLRO have to be computed for every specific model and cannot provide universal
classifications.

In the present work, we will show that macroscopic correlations are generally nonvanishing in generic symmetric states, and vanishing in symmetry-breaking
states. To this end, we consider the total correlations (classical plus quantum) between two generic subsystems $A$ and $B$ of the total system, as measured by the mutual information~\cite{NielsenChuang}:
\begin{equation}
\label{def1}
\mathcal I(A|B) = S(A) + S(B) - S(AB) \; ,
\end{equation}
where $S(X)$ is the von Neumann entropy of the density matrix pertaining to subsystem $X$, as shown pictorially in Fig.~\ref{system}. The
mutual information is indeed a {\em bona fide} measure of total correlations (classical plus quantum)~\cite{Olliver2001}. If for two arbitrary
subsystems $A,B$ spatially separated by arbitrarily large distances $d(A,B)$ the mutual information $\mathcal I(A|B)$ is vanishing, we are assured that
there are no macroscopic correlations and, in particular, no macroscopic entanglement and no macroscopic quantum correlations (all correlation and entanglement measures are non-negative defined). Otherwise, taking into account that the total system is in a global pure state, the two subsystems must be macroscopically entangled and quantum correlated. It is immediate to verify that $\mathcal I(A|B)$ vanishes on symmetry-breaking states that are product states of the form $|0\rangle^{\otimes N}$ and $|1\rangle^{\otimes N}$ , while for symmetric superpositions $\Psi_{\pm}$ it is always of order one, irrespective of the actual value of the distance $d(A,B)$.

We will show that the above result is in fact valid in general. In order to prove such statement, our strategy is the following. Starting from a factorization point, i.e. a point in which the system admits a fully separable pointer state as global ground state~\cite{Kurmann1982,GAI2008,GAI2009,GAI2010}, we will consider adiabatic deformations of the ground state and we will study the behavior of the
macroscopic mutual information. The deformation corresponds to the adiabatic continuation of the
ground state obtained by switching on the transverse magnetic field $h$. We will prove that, in the entire symmetry-breaking phase, the
total macroscopic correlations and, {\em a fortiori}, the macroscopic entanglement, as measured by the mutual information,
vanish in the maximally symmetry-breaking ground states at large spatial separations between arbitrarily selected macroscopic subsystems. 
On the other hand, we will also prove that, as long as the deformation is sufficiently small, the macroscopic mutual information remains finite in symmetric states. Later on, we will apply the results of the general analysis to specific spin models, showing that the result holding for slightly deformed symmetric states is in fact valid for all symmetric states in the entire symmetry-breaking (ordered) phase, until a quantum phase transition point is reached.

Proving the general result analytically would be quite a daunting task if one were to consider the  von Neumann entropy $S$. Rather, we will resort to a related quantity, the $2-$R\'enyi entropy, $S_2$, and the corresponding $2-$R\'enyi mutual information. For specific, computable, examples, we will then show that the general conclusions reached using the $2-$R\'enyi entropy hold as well using the von Neumann entropy.

The R\'enyi entropies of index $\alpha$ are defined as
\begin{equation}
S_\alpha (A) = (1-\alpha)^{-1}\log_2\mbox{Tr} \rho_A^\alpha \; ,
\end{equation}
and they are all {\em bona fide} entanglement monotones~\cite{Giampaolo2013}. In particular, the $2-$R\'enyi entropy $S_2$ is an experimentally accessible
quantity, since it is the expectation value of the swap operator on two copies of a quantum state~\cite{Ekert2002}, while the von Neumann
entropy $S$ is not an experimentally friendly observable, because it requires complete state tomography, which is essentially impossible on a many-body state.
Concrete proposals for measuring $S_2$ resort on quantum switches~\cite{Abanin2012}, or on multiparticle
interferometry~\cite{Zoller2012}. 

We thus consider the $2-$R\'enyi mutual information defined as
\begin{equation}
\mathcal I_2(A|B) := S_2(A) + S_2(B) - S_2(AB) \; .
\end{equation}
Unlike the quantum mutual information defined in Eq.~(\ref{def1}), the $2-$R\'enyi mutual information
may not be positive defined for every state in Hilbert space. In order to ensure positivity, one can regularize this definition in
terms of $2-$R\'enyi relative entropies, as shown in Refs.~\cite{renyimutual, renyimutual2}. However, on the class of states of interest for the present
investigation, non-regularized $2-$R\'enyi mutual information is always positive definite. Moreover, as already mentioned, in the
paradigmatic case of the exactly solvable one-dimensional quantum $XY$ model, we will also make direct comparison with the von
Neumann-based mutual information, finding complete agreement. Finally, the clustering property and other general properties of the $2-$R\'enyi mutual information that are at the core of our investigation and that we will prove below, can be proven without too much effort to hold valid also for the regularized versions. Therefore, in the following, also in order to avoid unnecessary formal mathematical complications, we will consider only the non-regularized version of the $2-$R\'enyi mutual information.

By adopting such strategy, we will obtain the following main result: in the ordered phase corresponding to
the spontaneous breaking of some symmetry of a many-body Hamiltonian with nonvanishing local order parameter, the mutual information
between two arbitrarily selected subsystems $A$ and $B$ separated by a distance $d(A,B)$ reaches its maximum in the symmetry-preserving ground
states and is upper bounded by $\exp(-O(d(A,B)))$ in the maximally symmetry-breaking ground states, i.e. the ground states that maximize
the order parameter. 

Therefore, we establish rigorously that - at least for some finite range of values within the phase -  spontaneous symmetry
breaking corresponds to the suppression of macroscopic coherent superpositions, and symmetry-breaking ground states are the ones
selected in real world by environmental decoherence, in complete analogy with pointer states. In the following, we specialize to the
class of global $\mathbb{Z}_2$ symmetry. In particular, we will consider the specific, but paradigmatic, example of the of quantum $XY$ 
models, and we will show that macroscopic entanglement survives until a quantum phase transition occurs. However, the central elements of 
this result hold in general and the remaining ones can be easily adapted to every instance of spontaneous symmetry breaking for different
Hamiltonians and different classes of symmetry groups.


\section{Clustering of 2-R\'enyi entropy and long-range order}

Macroscopic long-range total correlations are revealed by a nonvanishing mutual information
$\mathcal I$ between two arbitrary regions $A$ and $B$ separated by arbitrarily large distances. If $\mathcal I$ vanishes when $A$ and $B$
are separated by a distance larger than what defines the macroscopic interaction scale, then we are assured that there is no macroscopic
entanglement. 

Let us first consider the case of fully factorized ground states. These states are realized
at a precise and unique set of values of the Hamiltonian parameters in an ordered phase, the so-called factorization point or, in spin
systems, the factorizing field, first discovered in Ref.~\cite{Kurmann1982}. The general theory of ground-state factorization in terms of the response to local unitary perturbations was fully developed much later, in Refs.~\cite{GAI2008,GAI2009,GAI2010}. Factorized ground states can only occur in an ordered phase of strongly interacting many-body systems. Indeed, they are always maximally symmetry-breaking ground states with degeneracy equal to the dimension of the symmetry group and, being fully product states, they have a trivially vanishing mutual
information $\mathcal I$. Remarkably, at the factorization point, all other ground states can always be expressed as coherent linear
superpositions of the fully factorized and maximally symmetry-breaking ground states~\cite{Rossignoli2008}. Then, by construction,
symmetric ground states feature a nonvanishing $\mathcal I$, no matter how large the distance $d(A,B)$ between the $A$ and $B$ regions.
Moreover, there are no further types of entanglement and quantum correlations involved. Therefore, the classicality of symmetry-breaking states is immediately
verified at the factorization point.

As the Hamiltonian parameters are changed adiabatically and move away from the factorization point, ground-state entanglement does not
come solely from the macroscopic superposition of disconnected sectors, but also from the fact that fully factorized states are no
longer ground states. They can be represented as $U(s)|0\rangle^{\otimes N}, U(s)|1\rangle^{\otimes N}$, where $U(s)$ is the unitary
operator that connects the instantaneous ground states of the Hamiltonian $H(s)$. The operator $U(s)$ is an entangling
operator, showing that the symmetry breaking ground states are now entangled. This raises the problem whether they are still only
locally entangled or if they have developed some macroscopic entanglement. Similarly, for the symmetric states will be $U(s)\Psi_{\pm}$, it is
not immediately obvious to what extent their entanglement is macroscopic or not for a generic value of $s$. In the following, we show that the
entanglement and the correlations due to the adiabatic continuation $U(s)$ are local, i.e. they vanish in the limit of arbitrarily large
spatial separations $d(A,B)$.

In order to proceed, we need to discriminate clearly between the macroscopic and the local contributions to $\mathcal I$. We first recall that by $S_2(A)$ we mean the quantity $S_2(A)=-\log_2 \mathcal Q_A$ where
$\mathcal Q_A$ is the purity defined as $\mathcal Q_A= \mbox{Tr}[\rho_A^2]$, and $\rho_A$ is the reduced density matrix from the ground
state of the total system to the subsystem contained in the finite region $A$. To compute the R\'enyi entropy of order 2 we will use the
identity $\mathcal Q_A:=\mbox{Tr} \rho_A^2 = \mbox{Tr}(\mathcal{S}_A \rho^{\otimes 2})$, where $\rho^{\otimes 2}$ represents two copies
of the original full state on the doubled Hilbert space $\mathcal H\otimes \mathcal H'$. The operator $\mathcal S_A$ is the permutation
operator (swap operator) of order two with support on $\mathcal{H}_{AA'}$ only:
$\mathcal S_A = \tilde{\mathcal S}_A \otimes I_{\bar{A}}$ where $\tilde{\mathcal S}_A = \otimes_{a\in A} \tilde{\mathcal S}_a$ and
$\mathcal S_a$ is the permutation operator on the $a$-th spin of the system, i.e.
$\mathcal S_a |i_1,...i_a, ...,i_n\rangle\otimes  |j_1,...j_a, ...,j_n\rangle=  |i_1,...j_a, ...,i_n\rangle
\otimes  |j_1,...i_a, ...,j_n\rangle$.

Next, we exploit a locality result, stemming essentially from the Lieb-Robinson bound~\cite{lrb}. Indeed, for the mutual
information based on the von Neumann entropy, a recent seminal contribution has shown that $\mathcal I (A|B)$ is an upper bound for the
two-point correlation functions and a lower bound to exponentially decreasing functions of the ratio between the $d(A,B)$ and
the correlation length~\cite{VerstraeteMutualInfo}.

Following Hastings and Wen~\cite{Hastings2005}, let us consider a many-body Hamiltonian sum of local terms, $H(s)= \sum_i h_i(s)$ with a
finite gap $\Delta E$ above the low energy sector for some finite interval of values of the Hamiltonian parameters $s$ (then, out of
this interval a quantum phase transition may occur). Moreover, the local operators are assumed to be bounded: $\|h_i(s)\|<\infty$.
If the ground state of $H(s)$ is known for a particular, fixed set of values of the Hamiltonian parameters, say $s_0$, we may obtain it
for any other generic set $s$ by the quasi-adiabatic continuation $U(s)$ induced by a continuous deformation of $H(s)$. A local operator
$O_A$ with support on $A$ transforms as $O_A(s) = U^\dagger (s) O_A U(s)$. The new operator $O_A(s)$ has support on the whole Hilbert
space. Nevertheless, the locality result implies that we can arbitrarily approximate it with an operator $O_{A'}(s)$ that has support
only over the Hilbert space associated to a region with diameter \mbox{$diam(A')=diam (A) + l$}, as long as $l$ is larger than the
correlation length $\xi$ induced by the gap $\Delta E$, and by this making an error bounded in this way:
$\| O_A(s) - O_{A'}(s)\|\le K e^{-l/\xi}$. The constant $K$ grows like $l^D$ where $D$ is the spatial dimension (e.g. the lattice
dimension for localized spins).

Let now $\rho$ be the ground state of the system for $s=s_0$. The purity of the restriction of the evolved state $\rho (s)$ to a spatial
region $C$ reads
\begin{eqnarray}
\nonumber
\mathcal Q_C(s) &=& \mbox{Tr} \left[U(s)^{\otimes 2}\rho^{\otimes 2} (U^\dagger(s))^{\otimes 2} \mathcal S_C \right] \\
\nonumber
&=&  \mbox{Tr} \left[\rho^{\otimes 2} (U^\dagger(s))^{\otimes 2} \mathcal S_C U(s)^{\otimes 2}\right]\\
&\simeq& \mbox{Tr} \left[\rho^{\otimes 2}   \mathcal S_{C+l}  (s) \right] \; .
\end{eqnarray}
Here $\mathcal S_{C+l}$ denotes the permutation operator with support on spins that are at most at distance $l$ from $C$, and the
$\simeq$ sign means that the error is exponentially small in $l/\xi$. If the subsystem $C$ is the union of disjoint and macroscopically
separated subsets, $C = A \cup B$, we have
\begin{eqnarray}
\nonumber
\mathcal  Q_C &=& \mbox{Tr} \left[\rho^{\otimes 2} (U^\dagger(s))^{\otimes 2} \mathcal S_A \mathcal S_B U(s)^{\otimes 2}\right] \\
&=& \mbox{Tr} \left[\rho^{\otimes 2}  \mathcal S_{A+l}(s) \mathcal S_{B+l}(s) \right] \; .
\label{purity}
\end{eqnarray}
If the distance separating $A$ and $B$ is much larger than $l$, i.e. $ d(A,B)\gg l$, we can write
\begin{eqnarray}
\mathcal Q_C &\simeq& \mbox{Tr} \left[\rho^{\otimes 2}  \mathcal S_{A+l}(s) \otimes \mathcal S_{B+l}(s) \otimes I_E \right] \; ,
\end{eqnarray}
where $E$ is the complement to $A$ and $B$ together, see Fig.~\ref{system}.
Assume first that the initial state at $s=s_0$ is one of the completely factorized ground states (which, we recall, are also maximally
symmetry-breaking ground states). In this case,
$\rho(s_0) = \rho_A \otimes \rho_B \otimes \rho_E$ and we obtain
\begin{eqnarray}
\label{eq7}
\mathcal Q_C &\simeq& \mbox{Tr} \left[\rho^{\otimes 2}  \mathcal S_{A+l}(s) \right] \mbox{Tr} \left[\rho^{\otimes 2}
\mathcal S_{B+l}(s)  \right] \; .
\end{eqnarray}
Therefore, the purity in the region $C=A \cup B$ is the product of the purities in the two separated regions $A$ and $B$ from which it
follows immediately that
$\mathcal I_2(A|B)\simeq 0$.

Let us next consider the opposite case in which the ground state for $s=s_0$ is a macroscopic coherent superposition (Schr\"odinger cat)
of the fully factorized ground states. We show that any such superposition leads to a non vanishing $\mathcal I_2(A|B$. Since we need to
consider two copies of the ground state, we will label the factorized states by $|a\rangle$ and $|b\rangle$, respectively, on each copy.
So, we can write $|a\rangle = |a\rangle_A \otimes |a\rangle_B\otimes  |a\rangle_E$ and similarly for $|b\rangle$. Therefore,
$|\rho\rangle^{\otimes 2} = \sum_{a,b} \alpha_a \alpha_b |a\rangle_A |a\rangle_B |a\rangle_E \otimes |b\rangle_A |b\rangle_B
|b\rangle_E$, where $\sum_a |\alpha_a|^2=\sum_b|\alpha_b|^2=1$. For the sake of an explicit evaluation, consider the case of a doubly
degenerate ground state manifold and subsystems $A$ and $B$ with equal size (number of spins). For symmetric superpositions we then
obtain:
\begin{eqnarray}\nonumber
\!\! \!\!\!\!\mathcal I_2(A|B) \!&\!\!\simeq\!\!&  \log_2 4-2\log_2 \sum_{aba'b'}\langle a'b'|\mathcal S_{A+l}(s)|ab\rangle \\
\nonumber & \!\!+\!\!& \log_2 \sum_{aba'b'}\langle a'b'| \mathcal S_{A+l}(s)|ab\rangle^2  \\
&\!\!=\!\!&  \log_2 4 +\log_2\frac{\sum_{aba'b'}\langle a'b'|\mathcal S_{A+l}(s)|ab\rangle^2}
{(\sum_{aba'b'} \langle a'b' | \mathcal S_{A+l}(s) | ab \rangle )^2} \label{main} \, .
\label{mainresult}
\end{eqnarray}
Since for $s=0$ the expectation values of $\langle a'b'| \mathcal S_{A}(0)|ab\rangle$ are positive definite, by continuity they are
still positive for a small enough value of $s$. So we see that, for small $s$, the second term in Eq.~\ref{mainresult}, is negative but
that the $\mathcal I_2(A|B)$ stays strictly positive, for arbitrary $d(A,B)$. This is also true for any non trivial superposition of the
symmetry breaking sectors given by the amplitudes $\alpha_a$, so that only the maximally symmetric breaking ones have exactly zero
$\mathcal I_2(A|B)$ (as $d(A,B)\rightarrow\infty$).

In the following, we determine the exact value of  $\mathcal I_2(A|B)$ for macroscopic coherent superpositions with arbitrarily large
$d(A,B)$ in the entire ordered phase, beyond the perturbative case of small $s$ in the case in which $A$ and $B$ are made by a single
spin. We  prove this result explicitly for models with $\mathbb{Z}_2$ symmetry, but the central elements of the proof are valid for
arbitrary symmetry groups and arbitrary dimension of $A$ and $B$. In fact, our proof implies that for a maximally
symmetry broken ground state all the correlations function between two very far subsystems factorize in the product of the expectation
value of the local operators. And this implies that the mutual information between $A$ and $B$ must vanish.
When we turn to consider symmetric ground states, some of these local expectation values must vanish because the local operator does not
commute with at least one of the parity operators that define the symmetry group of the Hamiltonian. As result the mutual information
is expected to be different from zero. A comprehensive analysis on the dependence on the size of the subsystems and/or the
symmetry group of the Hamiltonian is in progress~\cite{inprogress}.

\section{Long-range mutual information in models with $\mathbb{Z}_2$ symmetry}

In the following, we focus on spin-$1/2$ systems with a global $\mathbb{Z}_2$ symmetry, thus described by Hamiltonians that commute with
the parity operator along a fixed spin direction, i.e. \mbox{$\mathbb{P}_\mu=\otimes_i \sigma_i^\mu$}. In such systems spontaneous
symmetry breaking is associated to the presence of a two-fold degenerate ground state and an off diagonal long-range order along a spin
direction $\sigma_i^\nu$ that is orthogonal to $\sigma_i^\mu$. We  show that throughout the entire ordered phase the long-range mutual
information vanishes identically on states that maximally break the symmetry, while it remains strictly positive on any macroscopic
coherent superposition of the two broken symmetry sectors. For the sake of simplicity we focus on the case in which subsystems $A$ and
$B$ are each one made by a single spin but the results can be extended straightforwardly to more general choices.

The two orthogonal symmetric ground states $|e\rangle$ and $|o\rangle$ (respectively, the even-symmetric and the odd-symmetric ground
states) form a convenient basis that allows to write all other ground states $|g\rangle$ in the ordered phase as their linear
superpositions: $|g(u,v)\rangle =u |e\rangle + v |o\rangle$. The reduced density matrix $\rho_C$ of the two-spin block $C = A \cup B$
can be expressed in terms of the  $2$-point correlation functions as follows~\cite{Osborne2002}:
\begin{equation}
\label{eq:defreduce}
 \rho_{C}\!(u,v) = \! \frac{1}{4} \sum_{i_A,i_B}  G^{i_A,i_B}(u,v) \sigma_{A}^{i_A}  \sigma_{B}^{i_B}  \; ,
\end{equation}
where the expectations $G^{i_A,i_B}(u,v)$=$ \langle g(u,v)\!| \sigma_{A}^{i_A}\!  \sigma_{B}^{i_B} \! |g(u,v)\!\rangle$ are on products
of Pauli matrices $\sigma_{A}^{i_A}$ and $\sigma_{B}^{i_B}$. As shown in Ref.~\cite{Cianciaruso2014}, all correlation functions can be
associated to spin operators that either commute or anti-commute with the parity operator $\mathbb{P}_\mu$. Therefore, the reduced
density matrix $\rho_C(u,v)$ can be expressed as the sum of a symmetric part that coincides with the density matrix of the symmetric
ground state, $\rho_C^{(s)}(u,v)$, and an antisymmetric part, that is a traceless matrix, $\rho_C^{(a)}(u,v)$. Taking into account the
fact that the two symmetric ground states fall in two orthogonal eigenspaces of the Parity, it is straightforward to verify that
$\rho_C^{(s)}(u,v)$ is independent of the superposition amplitudes and hence $\rho_C^{(s)}(u,v) \equiv \rho_C^{(s)}$.

The reduced density matrix of a symmetric ground state thus reads
\begin{equation}
\label{eq:defreduce1}
\! \! \rho_{C}^{(s)}= \!\! \frac{1}{4} (1\!\!1+\!m_\mu(\sigma_{A}^{\mu}+\sigma_{B}^{\mu})\!+\!m_\mu^2 \sigma_{A}^{\mu}\sigma_{B}^{\mu}+
 \!m_\nu^2 \sigma_{A}^{\nu}\sigma_{B}^{\nu}) \; .
\end{equation}

In Eq.~(\ref{eq:defreduce1}) $m_\mu$ is the expectation value of the local operator that commutes with the parity while $m_\nu$ is the
local order parameter. Exploiting Eq.~(\ref{eq:defreduce1}) one can derive the mutual information $\mathcal{I}_2^{(s)}$ and evaluate
its asymptotic expression for $d(A,B) \rightarrow \infty$:
\begin{equation}
\mathcal{I}_2^{(s)}(\infty)=\log_2 \left[1+\frac{m_\nu^4}{(1+m_\mu^2)^2}\right] \; .
\end{equation}

As $m_\nu \neq 0$ throughout the entire ordered phase, the above relation shows the presence of macroscopic entanglement and
correlations in the symmetric, coherent superposition ground states throughout the entire phase. Up to this point, this result is valid
for any model with $\mathbb{Z}_2$ symmetry. The actual values of  $m_\nu$ of course depend on the specific model considered. Here, we
analytically compute the result for the quantum $XY$ model.

The one-dimensional spin-$1/2$ quantum $XY$ Hamiltonian with ferromagnetic nearest-neighbor interactions in a transverse field with
periodic boundary conditions reads:
\begin{equation}\label{eq:XYmodelhamiltonian}
H \! =\!-\!\sum_{i=1}^{N}\! \! \left[\!\left(\!\frac{1+\gamma }{2}\!\right) \! \sigma_i^x \sigma_{i+1}^x \!+\!
\left(\!\frac{1-\gamma }{2}\!\right)\!\sigma_i^y
\sigma_{i+1}^y \!+\! h \sigma_i^z\right]\! \; ,
\end{equation}
where $\sigma_i^\mu$, $\mu = x, y, z$, are the Pauli spin-$1/2$ operators acting on site $i$, $\gamma$ is the
anisotropy parameter in the $xy$ plane, $h$ is the transverse magnetic field along the $z$ direction, and the periodic boundary
conditions $\sigma_{N+1}^\mu \!\equiv\! \sigma_1^\mu$ ensure invariance of the model Hamiltonian under spatial translations.

\begin{figure}[t]
\includegraphics[width=6cm]{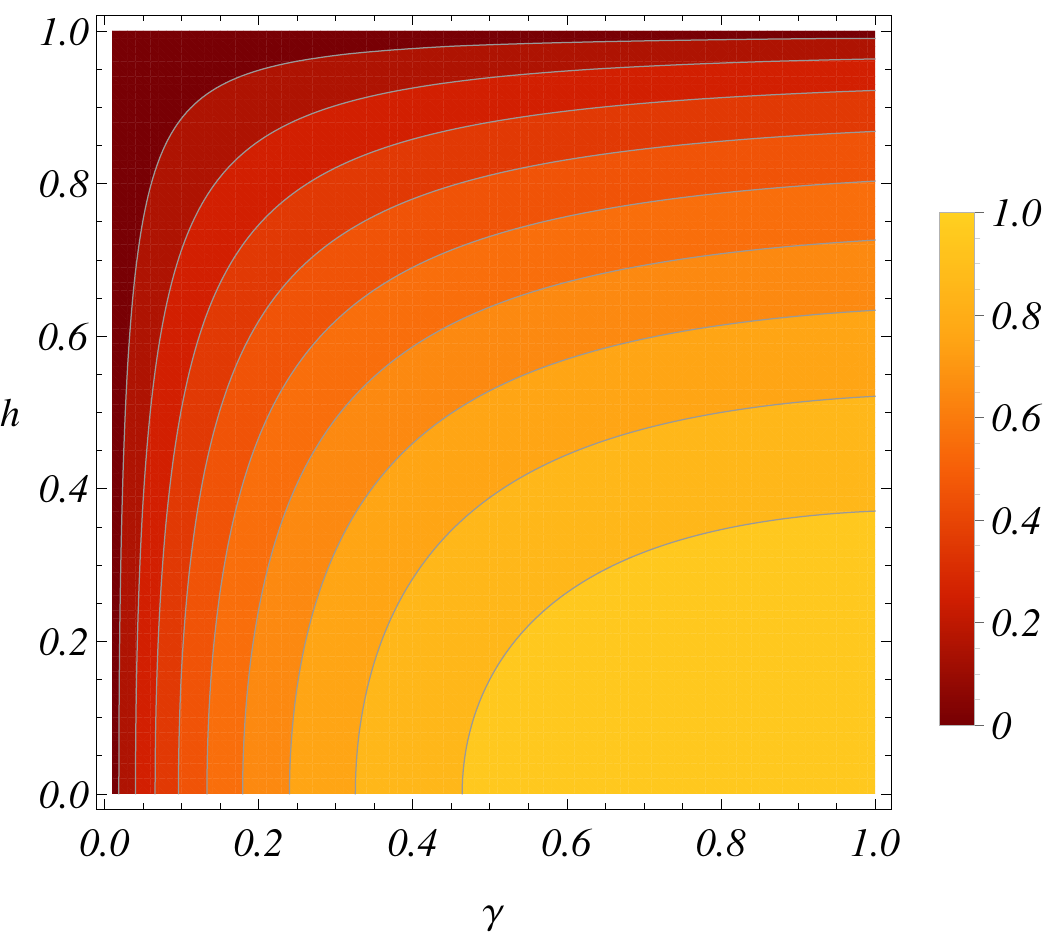}
\caption{(Color Online) Behavior of the mutual information $(\mathcal{I}_2^{(s)}(\infty))$ between two spins at infinite distance in
the symmetric ground state of the one-dimensional $XY$ model (thermodynamic limit) as a function of the anisotropy $\gamma$ and
transverse field $h$ in the ferromagnetic phase $0 \leq h \leq h_c = 1$. Within the ferromagnetic phase, it is always $m_\nu > 0$, and
hence $\mathcal{I}_2^{(s)}(\infty) > 0$. On the other hand, $m_\nu = 0$ either at $h_c=1$, or at $\gamma=0$. Only at these points
$\mathcal{I}_2^{(s)}(\infty) = 0$.}
\label{symmetryvalue}
\end{figure}

Such model can be solved analytically~\cite{LSM1961,BMD1970,BM1971} and, hence, the phase diagram can be determined exactly and in great
detail. In the thermodynamic limit, for any $\gamma\!\in\!(0,1]$, a quantum phase transition occurs at the critical value $h_c = 1$ of
the transverse field. For $h\! <\! h_c\!=\!1$ the system is ferromagnetically ordered and is characterized by a twofold ground-state
degeneracy such that the $\mathbb{Z}_2$ parity symmetry under inversions along the spin-$z$ direction is broken by some elements of the
ground space. Using the analytical solution, in Fig.~\ref{symmetryvalue} we have plotted the behavior of
$\mathcal{I}_2^{(s)}(\infty)$ in the ferromagnetic phase.

Given the two symmetric ground states, the so-called even $|e\rangle$ and odd $|o\rangle$ states belonging to the two
orthogonal subspaces associated to the two possible distinct eigenvalues of the parity operator, any symmetry-breaking linear
superposition of the form
\begin{equation}\label{eq:groundstates}
|g(u,v)\rangle = u |e\rangle + v |o\rangle \;
\end{equation}
is also an admissible ground state, with the complex superposition amplitudes $u$ and $v$ constrained by the normalization condition
\mbox{$|u|^2\!+\!|v|^2\!=\!1$}. Taking into account that the even and odd ground states are orthogonal, the expectation values of
operators that commute with the parity operator are independent of the superposition amplitudes $u$ and $v$. On the other hand, spin
operators that do not commute with the parity may have nonvanishing expectation values on such linear combinations and hence break the
symmetry of the Hamiltonian Eq.~(\ref{eq:XYmodelhamiltonian}).

In the asymptotic macroscopic regime $d(A,B)\rightarrow\infty$, the general two-spin reduced density matrix for an arbitrary ground
state reads
\begin{eqnarray}
\label{eq:defreduce2}
\! \rho_{C}\!(u,v)\!\! &\!=\!& \!\!\rho_{C}^{(s)}+ \frac{1}{4} (u v^*+v u^*)\left[ m_\nu(\sigma_{A}^{\nu}+\sigma_{B}^{\nu})
\right.\nonumber \\
 &+& \left. m_\mu m_\nu( \sigma_{A}^{\mu}\sigma_{B}^{\nu}+
 \sigma_{A}^{\nu}\sigma_{B}^{\nu}) \right] \; .
\end{eqnarray}
The corresponding expression for the mutual information $\mathcal{I}_2(\infty)$ reads
\begin{equation}
\label{mutualgeneric}
 \!\!\mathcal{I}_2(\infty)\!=\!\log_2\!\! \left[1+\frac{m_\nu^4(1-(uv^*+vu^*)^4)}{(1+m_\mu^2+(uv^*+vu^*)^2m_\nu^2)^2} \right] \; .
\end{equation}

\begin{figure}[t]
\includegraphics[width=8.5cm]{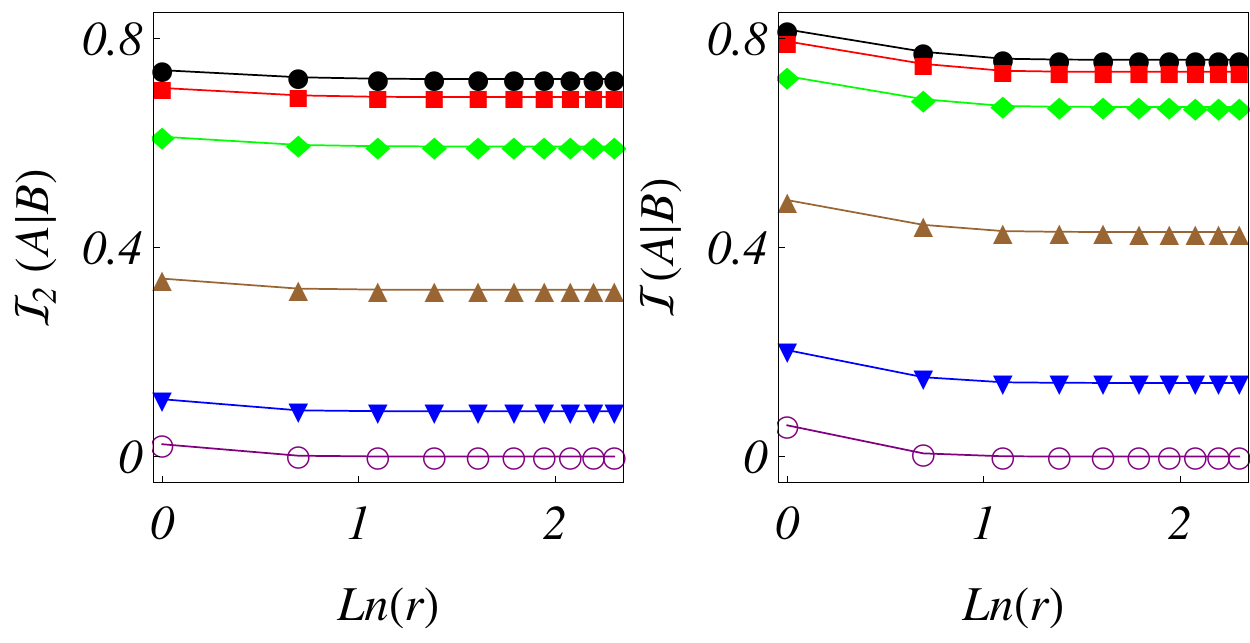}
\caption{(Color online) Behavior of the mutual information based on the $2$-R\'enyi entropy, $\mathcal{I}_2(A|B)$ (left), and the von
Neumann entropy, $\mathcal{I}(A|B)$ (right), as functions of the logarithm of the inter-spin distance $r$ for different superpositions
in the ferromagnetic phase of the one-dimensional $XY$ model at $(\gamma=h=0.5)$. Black circles: $u=1$, $v=0$ (symmetric state);
red squares: $u=\cos(0.05 \pi)$, $v=\sin(0.05 \pi)$; green diamonds: $u=\cos(0.1 \pi)$, $v=\sin(0.1 \pi)$; brown up-triangles:
$u=\cos(0.15 \pi)$, $v=\sin(0.15 \pi)$; blue down-triangles: $u=\cos(0.2 \pi)$, $v=\sin(0.2 \pi)$; violet empty circles:
$u=\cos(0.25 \pi)$, $v=\sin(0.25 \pi)$ (maximally symmetry-breaking ground state). The two definitions feature the same qualitative
behavior; in particular, they both vanish in and only in the maximally symmetry-breaking ground states.}
\label{funcr}
\end{figure}

\begin{figure}[t]
\includegraphics[width=8.5cm]{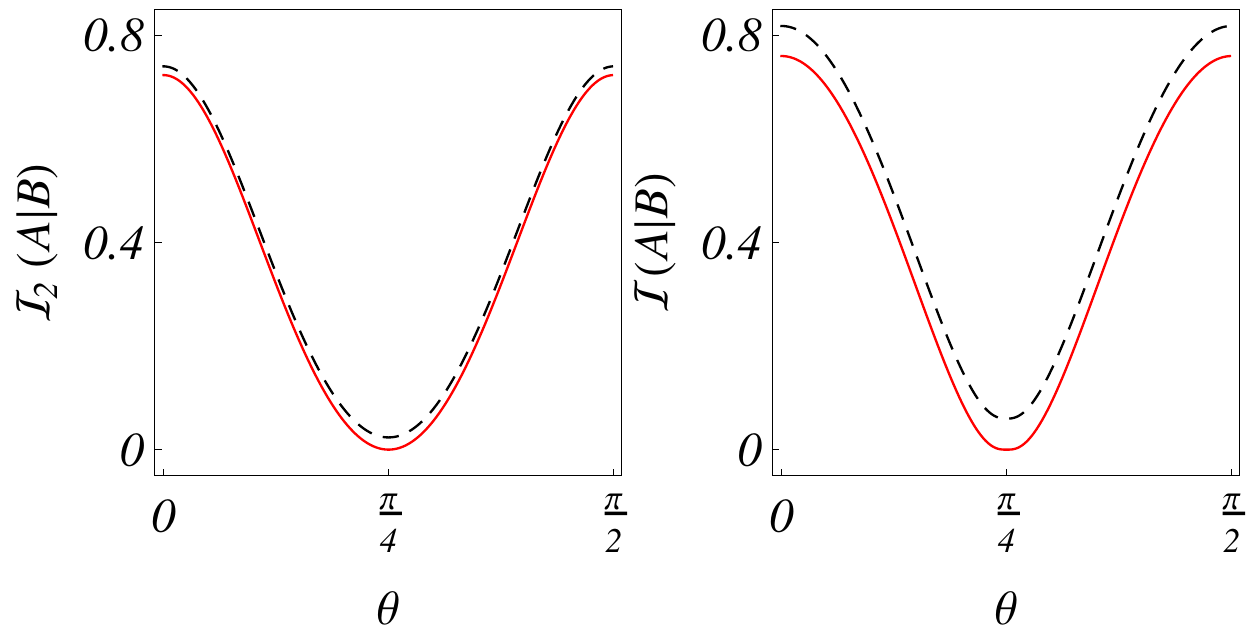}
\caption{(Color online) Behavior of the mutual information based on the $2$-R\'enyi entropy, $\mathcal{I}_2(A|B)$ (left), and the von
Neumann entropy, $\mathcal{I}(A|B)$ (right), as functions of the ground-state superposition parameter $\theta$, with the parametrization
$u = \cos \theta$, $v = \sin \theta$, for two extreme values of the distance $r$ between $A$ and $B$. Dashed black curve: $r = 1$.
olid red curve: $r = \infty$. The absolute minimum is always realized at $\theta = \pi/4$, corresponding to the maximally
symmetry-breaking ground state. At this point, both measures of mutual information vanish exactly for $r = \infty$.}
\label{functheta}
\end{figure}

Due to the normalization constraint, $|u|^2+|v|^2=1$, the fraction in Eq.~(\ref{mutualgeneric}) is semi-positive defined and vanishes
only either at $m_\nu=0$, i.e. in the disordered, classical paramagnetic phase, or when $(uv^*+vu^*)=1$. Therefore, in the ordered
phase the only ground states with vanishing long-range mutual information, and hence vanishing macroscopic entanglement and
correlations, are the maximally symmetry-breaking ground states. At the other end of the spectrum, it is easy to verify that the maximum
of $\mathcal{I}_2$, for a fixed value of the parameters $m_\mu$ and $m_\nu$, is always achieved in the totally symmetric (even) and
antisymmetric (odd) states, the absolute maximum being obtained for $m_\mu=0$ and $m_\nu=1$. Finally, since the one-dimensional $XY$
model allows for the exact evaluation of all entropies in the R\'enyi hierarchy, in Fig.~\ref{funcr} we compare the mutual information
based on the $2$-R\'enyi and the one based on the von Neumann entropy, finding complete qualitative and quantitative agreement. In
particular, they both vanish in, and only in, the maximally symmetry-breaking ground states. Finally, using the parametrization
$u = \cos \theta$, $v = \sin \theta$, in Fig.~\ref{functheta} we also report and compare for completeness the behavior of the
$2$-R\'enyi based and the von Neumann based mutual information as functions of the superposition parameter $\theta$ for the two
extreme cases of nearest-neighbor distance $r=1$ and asymptotic distance $r=\infty$, finding again perfect agreement.

\section{Comparison with other indicators of macroscopic coherent superpositions}

Quantum discord~\cite{Zurek2000,Olliver2001}
is a measure of quantum correlations more general than entanglement that may exist in mixed quantum states, including
separable ones. It is defined as the difference between mutual information -which accounts for all correlations, both classical and
quantum- and the optimal classical correlations between $A$ and $B$, by maximizing
over all the measurement on $B$:
$
  {\cal C}_{AB} = \max_{\{\hat{B}_k\}} [ {\cal S}(\hat{\rho}_A) - {\cal S_C} (\hat{\rho}_{AB} | \{\hat{B}_k\}) ]
$.
It is therefore sensible to verify whether it may be a good quantifier of macroscopic quantum coherence. The long-range
pairwise quantum discord between two spins in the ground state of the one-dimensional $XY$ chain has been recently investigated in
Refs.~\cite{TRHA2013,Cianciaruso2014,Cianciaruso2014-1,Campbell2013}.
It turns out that such quantity features a long-range behavior quite analogous to that of the
mutual information, with the crucial difference that it vanishes identically in all possible ground states as $m_\mu \rightarrow 0$.
From a mathematical point of view this can be easily explained considering that, in such a case, the two-spin reduced density matrix in
the symmetric ground states at asymptotically large inter-spin distance is indistinguishable from the one obtained by the
symmetry-breaking Gibbs states at zero temperature.

The localizable entanglement in a many-body pure state is defined as the maximal amount of pairwise entanglement between two spins $i,j$
at arbitrary distance that can be achieved, on average, by performing generalized measurements on all other
spins~\cite{Localizable1,Localizable2}. This naturally defines an entanglement length $\chi$ that diverges in symmetric states.
Numerics show that in the maximally symmetry-breaking ground states of the $XY$ chain the localizable entanglement behaves like the
connected spin-spin correlation functions $Q_{xx}$ that are bound to decay exponentially at long distance. Therefore long-range pairwise
entanglement defined via the localizable entanglement features a behavior quite similar to that of the mutual information. However,
this is just pairwise entanglement (although at long distance), so it does not quite capture the notion of a macroscopic superposition.
Moreover, the mutual information is more readily generalized to thermal mixed states, the Gibbs states at finite temperature, for
which the evaluation of the mutual information presents no particular difficulty.


\section{Conclusions and outlook}

We investigated  macroscopic entanglement~\cite{vedralnature} through the behavior of the quantum mutual information between two
macroscopically separated blocks of dynamical variables in the ground state of many-body systems featuring spontaneous symmetry
breaking. This quantity detects macroscopic total correlations, including entanglement. The main result of this paper is that in the
{\em entire phase with broken symmetry}, the symmetry-breaking states have vanishing long-distance mutual information, while the latter
remains finite for any non maximally symmetry-breaking superposition, attaining a maximum for the totally symmetric states. This fact is
easy to prove when considering symmetry-breaking states that are completely unentangled (fully factorized), whose symmetric
superpositions are GHZ states. In order to prove this feature in the entire ordered phase, a much more challenging task, we followed a
strategy based on two ingredients: (i) adopting measures of mutual information based on the $2-$R\'enyi and on the von Neumann
entropies, and (ii) exploiting locality results about quasi adiabatic continuation of quantum states derived by using the
Lieb-Robinson bounds~\cite{lrb, hastingslrb, hastingslrb2}. In this way we were able to prove that spontaneous symmetry breaking selects
the many-body states with vanishing long-distance mutual information, and thus macroscopically least entangled, and therefore most
classical.

In perspective, we are concerned with the investigation of several open problems. In particular, it would be interesting to extend our
analysis to the case of subsystems of arbitrarily variable size, in order to observe possible threshold effects, and to generic classes
of symmetry groups~\cite{inprogress}. Moreover, we are interested in studying the case of globally mixed states~\cite{vedralnjp} and, in particular,
equilibrium thermal states of models featuring spontaneous symmetry breaking below a critical temperature. At thermal equilibrium the system will be described by the Gibbs state
\mbox{$\rho_{eq} = Z^{-1}\sum_{i,a}
e^{-\beta E_i}|E_i\rangle\langle E_i|_a$}, where with $a$ we have explicitly labeled different sectors. Below a critical temperature
$T_c$, if $a$ labels the sectors with broken symmetry, spontaneous symmetry breaking means that in every single realization the label
$a$ will be fixed by the initial conditions. Therefore, our statement is that Gibbs states obtained by fixing $a$ feature the least
long-range mutual information compared to all other non-maximally symmetry-breaking Gibbs states~\cite{inprogress}.

From a different perspective, we remark that many-body localization has recently become a subject of great interest for the condensed
matter community. Systems with strong disorder featuring many-body localization fail to thermalize and to obey the eigenstate
thermalization hypothesis~\cite{husereview}, because their eigenstates are more weakly entangled than in typical non-integrable systems.
It would be interesting to see how the techniques developed in the present work might help in providing rigorous results regarding the
clustering of mutual information in such systems. In fact, our techniques can be used to investigate also systems that involve
long-range entanglement and correlations, such as topologically ordered states~\cite{wen,wenzeng} and their resilience in the presence
of perturbations or at finite temperature~\cite{topent1, topent2, topent3}.



{\em Acknowledgments.---} We acknowledge useful conversations with F. Verstraete. We also acknowledge useful comments by G. Adesso and
P. Zanardi on an earlier draft of the manuscript. AH thanks the National Basic Research Program of China Grant 2011CBA00300, 2011CBA00301 the National Natural Science Foundation of China Grant  No. 61361136003 and No. 11574176. S.M.G. and F.I. acknowledge the
FP7 Cooperation STREP Project EQuaM - Emulators of Quantum Frustrated Magnetism (GA No. 323714).

\end{document}